# Towers on the Peaks of Eternal Light: Quantifying the Available Solar Power

Amia Ross[1], Sephora Ruppert[1], Philipp Gläser[2], and Martin Elvis[3]

[1]Harvard College; [3]Harvard-Smithsonian Center for Astrophysics, Harvard University, Cambridge, MA, USA

[2]Department of Planetary Geodesy, Technische Universität Berlin, Berlin, Germany

**Corresponding Author:**

Amia Ross

In Care Of: Dr. Martin Elvis

Harvard-Smithsonian Center for Astrophysics

60 Garden Street

Cambridge, MA, 02138

Tel: 847-877-2699

Email: amia_ross@college.harvard.edu

**Running Title:** Available Solar Power on PELs

# ABSTRACT

The Peaks of Eternal Light (PELs), that are largely unshaded regions mostly at the lunar south pole, have been suggested as a source of solar power for mining the water and other volatiles in the nearby permanently dark regions. As mining is a power-intensive activity, it is interesting to estimate the maximum solar power that could be generated at the PELs. Here we use average percentage illumination maps for a range of heights above the local topography from 2 m to 2 km to determine the total power available as a function of time of lunar day. Overshadowing of highly illuminated areas by towers placed in sunward locations (at a given time of day) limits the total power to much smaller values than the highly illuminated area would suggest. We find that for near-term realizable towers (up to 20 m), the upper limit to the time-averaged power available is ~55 MW at >70% illumination, and ~6 MW at >90% illumination. For the more distant future a maximum time-averaged power of order 21000 MW at >70% illumination could be realizable for towers up to 2 km in height, and ~5270 MW, respectively, at 90% illumination. Towers 1 km high provide about a factor 2.7 times less power. The variation with lunar time of day ranges from a factor of 1.1 to ~ 3.

**KEYWORDS:** moon, lunar mining, lunar tower, lunar solar power, resource management

## 1. Introduction

The so-called "Peaks of Eternal Light" [1] near the north and south poles of the Moon are almost continuously in sunlight [2]. It is the small tilt of the Moon's axis to the Ecliptic plane (1.5º versus 23.5º for the Earth) [3] that makes this illumination possible [4]. As a result, the PELs offer the prospect of generating near continuous solar power. At the south pole in particular, these peaks lie within a few kilometers of substantial water deposits in the permanently dark regions [5]. This power is then attractive for mining purposes. Mining is, however, a power-intensive activity, with estimated power requirements from a few megawatts [6] to several gigawatts [7]. It is important, then, to estimate how much solar power could be generated at the PELs, and how much the power availability would vary during a lunar day. The total area at ground level is small, of the order a few km² [8]. This suggests quite a limited maximum power.

However, Gläser et al. [8] also showed that the percentage of time with solar illumination at the PELs grew substantially with even a modest elevation of 2 meters.

In this paper we consider the total area available for solar power generation for different height towers. This total area depends on the orientation of the Sun to the local topography. During a single 27.3 day long lunar day the Sun circles near the horizon as seen from the PELs. As a result, much of the area will be shadowed if solar towers are built in front of them. The total power will then depend on the projected length that is illuminated rather than the total area. This topography is not symmetric, so different lengths of high illumination ridges are presented to the Sun throughout the lunar day due to projection effects. Hence, we also considered the variation of the total available power through a lunar day.

We considered a wide range of elevations. First, we investigate modest heights that might be achieved in the near future (2 - 20 m). For example, the towers considered for the planned ESA Lunar Village [9] could be installed at the PELs within a decade. These towers have a height of 15 meters. In the longer term, given that the Moon lacks an atmosphere, is seismically quiet, and has ~⅙ the surface gravity of the Earth, much taller towers can, in principle, be constructed on the Moon, potentially using local resources [10]. We therefore also consider truly tall towers, from 100 m up to 2000 m in height.

## 2. Data

High resolution maps of the average percentage solar illumination and intensity were generated for a 50 by 50 km region centered on the lunar south pole. This region covers all of Shackleton crater and the ridge between Shackleton and de Gerlache craters, referred to as "connecting ridge" [11]. The region does not include Malapert Mountain. The maps were constructed based on a digital terrain model (DTM) [8] constructed from laser tracks of the Lunar Orbiter Laser Altimeter (LOLA) [12] instrument on the NASA Lunar Reconnaissance Orbiter [13], following the procedure of Gläser et al. [8]. Illumination and intensity values were derived every two hours and averaged over a 20-year period (January 01, 2020 to January 01, 2040) in order to span the 18.6-years lunar precessional cycle [8]. Mean percentage illumination values were calculated at

elevations of 0, 2, 10, 16, 20, 100, 500, 1000, 1500, and 2000 meters above the surface. The maps have 20 x 20 m pixels and so form a 2500 x 2500 array centered at (latitude = -90°, longitude = 0°) in gnomonic map coordinates (using a lunar radius 1737.4 km) [14]. Figures 1 and 2 show the resulting maps in order of increasing elevation. Regions in red or dark red have interestingly high values of average solar illumination, i.e. >80%. It is clear that for towers up to 20m in height these regions are much smaller than kilometer-scale towers (100 m - 2000 m), for which far larger areas have high solar illumination (Figure 2).

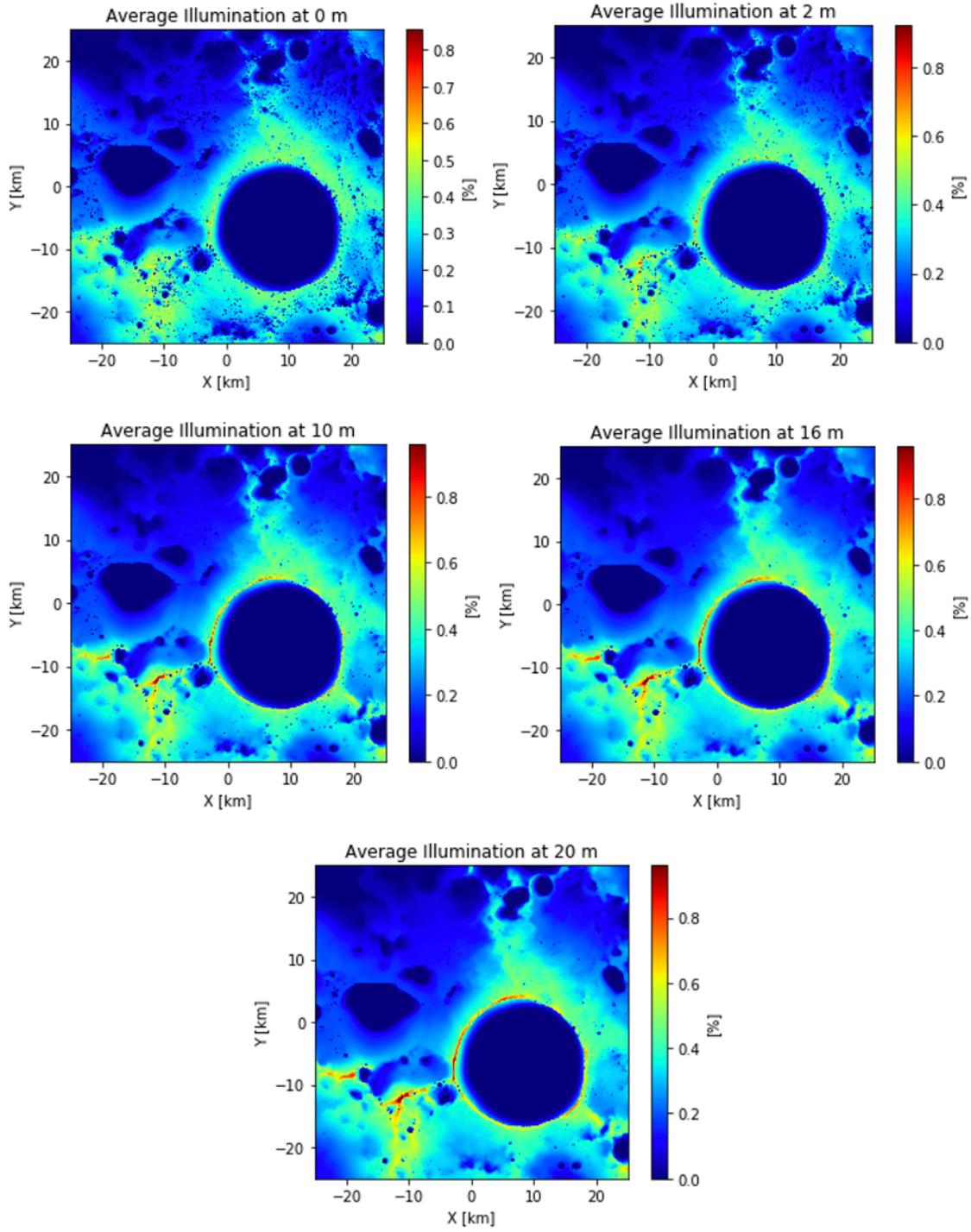

Fig. 1: Average Illumination at 0, 2, 10, 16, and 20m, X and Y coordinates in km.

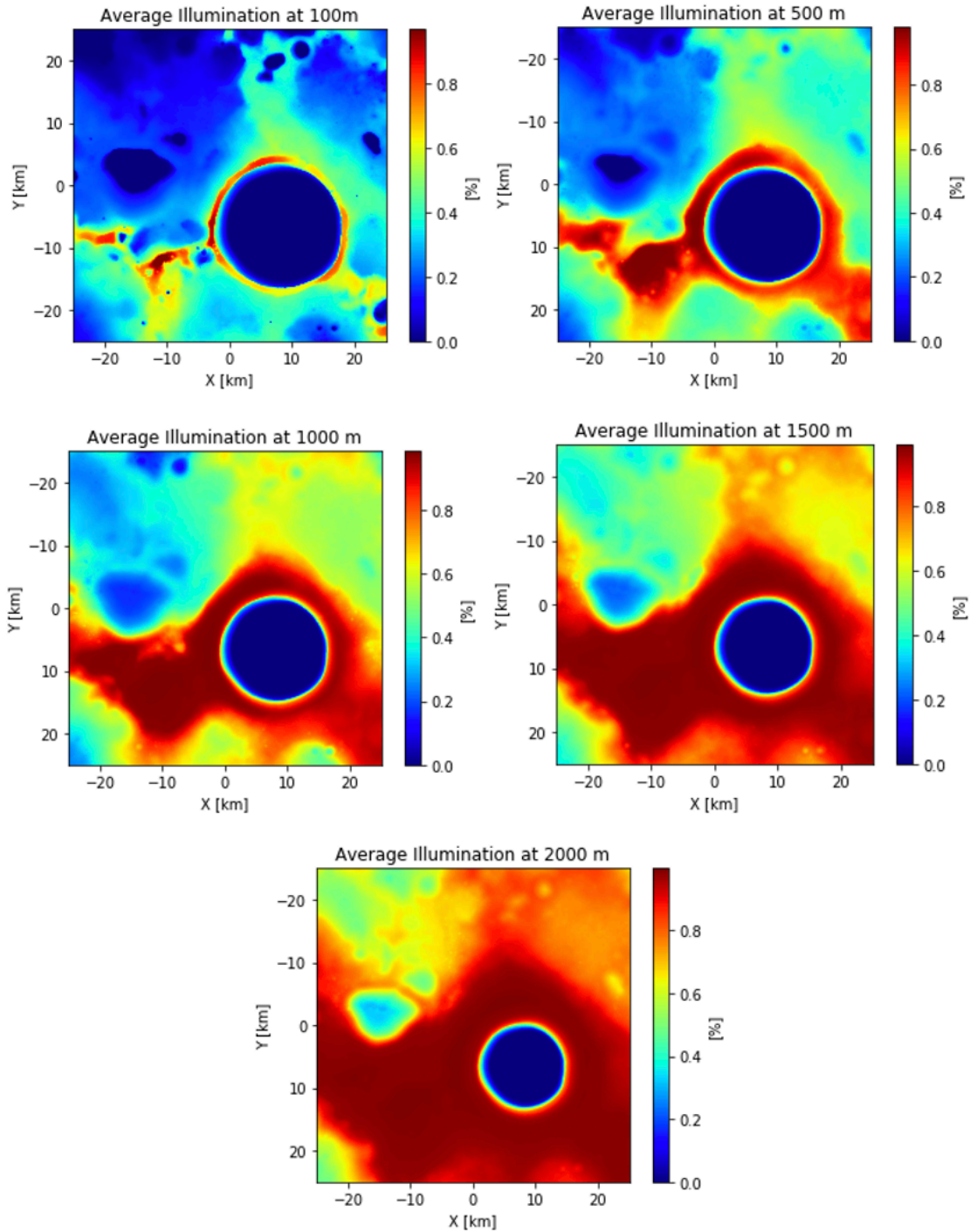

Fig. 2: Average Illumination at 100, 500, 1000, 1500, and 2000m, X and Y coordinates in km.

## 3. Data Analysis

### 3.1 Area versus Illumination threshold percentage

The high illumination area shrinks as a function of the threshold illumination value chosen. Figure 3 shows this trend for a 100 m elevation.

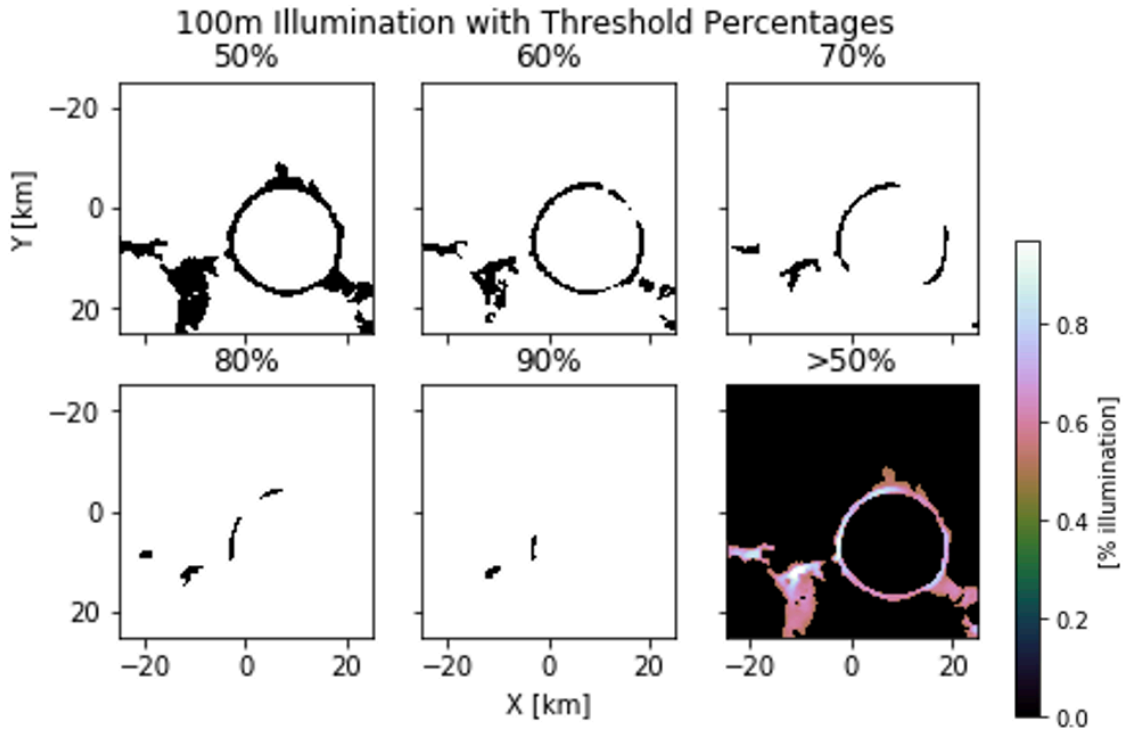

Fig. 3: Illumination Threshold Maps at 100m, bottom right with illumination 50% and higher in 10% steps

More quantitatively, Figure 4 (left) shows the total area as a function of the illumination threshold against illumination percentages between 50-100%. To see the area available at the highest illumination values Figure 4 (right) shows a zoom into this part of the plot. Only for >100 m elevation is more than 2 sq. km illuminated for over 90% of the time.

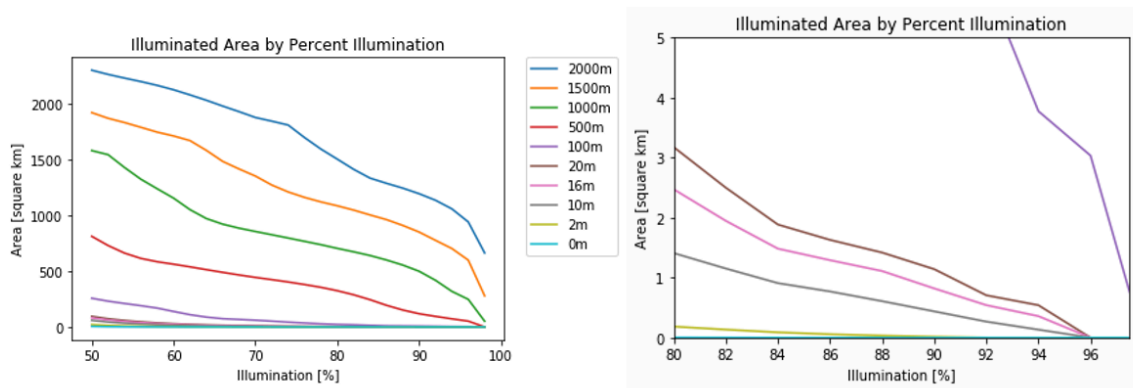

Fig. 4: Illuminated Area by Illumination threshold percentage. Left: 50% - 100%, right: >80% illumination

Repeating this exercise for the other elevations gives the illuminated area against elevation for different illumination thresholds (Figure 5, solid points). Illuminated area per elevation and illumination threshold is given in Table 1.

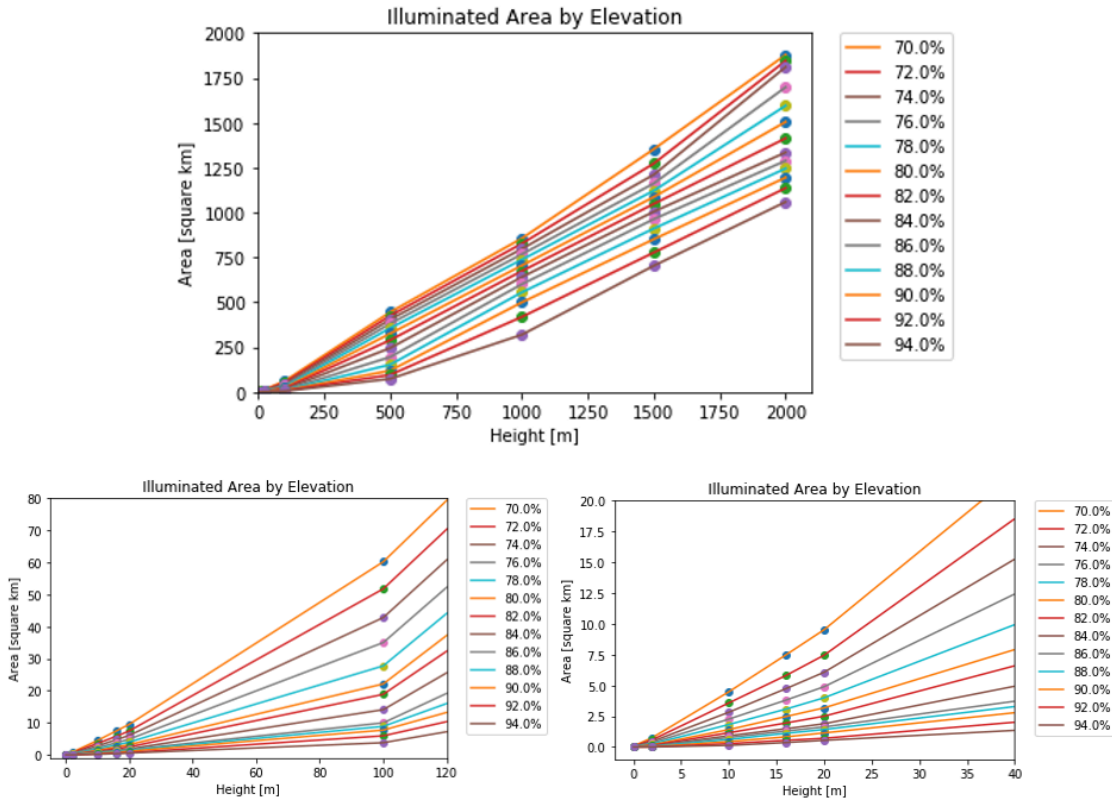

Fig 5: Illuminated Area by Elevation (upper), with blowup panels for heights <= 100m (lower left) and 20m (lower right).

Table 1: Illuminated area in square km by % illumination and height.

| Height [m] | 0 | 2 | 10 | 16 | 20 | 100 | 500 | 1000 | 1500 | 2000 |
|---|---|---|---|---|---|---|---|---|---|---|
| % Illum | - | - | - | - | - | - | - | - | - | - |
| 70 | 0.0724 | 0.767 | 4.48 | 7.48 | 9.52 | 60.2 | 444 | 856 | 1350 | 1880 |
| 72 | 0.048 | 0.590 | 3.58 | 5.81 | 7.45 | 51.8 | 424 | 827 | 1270 | 1850 |
| 74 | 0.0308 | 0.458 | 2.82 | 4.74 | 6.07 | 42.8 | 403 | 797 | 1210 | 1810 |
| 76 | 0.0148 | 0.349 | 2.26 | 3.81 | 4.89 | 35.0 | 380 | 767 | 1160 | 1700 |
| 78 | 0.0072 | 0.254 | 1.82 | 3.08 | 4.00 | 27.8 | 355 | 737 | 1120 | 1590 |
| 80 | 0.0036 | 0.185 | 1.41 | 2.47 | 3.18 | 22.1 | 325 | 704 | 1090 | 1500 |
| 82 | 0.002 | 0.134 | 1.15 | 1.95 | 2.50 | 18.9 | 288 | 674 | 1050 | 1410 |
| 84 | 0.0004 | 0.0884 | 0.909 | 1.48 | 1.89 | 14.1 | 244 | 640 | 1000 | 1330 |
| 86 | 0 | 0.0564 | 0.770 | 1.29 | 1.63 | 9.97 | 194 | 600 | 963 | 1290 |
| 88 | 0 | 0.0336 | 0.608 | 1.11 | 1.42 | 8.88 | 152 | 555 | 910 | 1240 |
| 90 | 0 | 0.0152 | 0.438 | 0.820 | 1.14 | 7.71 | 118 | 499 | 851 | 1190 |
| 92 | 0 | 0.0008 | 0.267 | 0.542 | 0.706 | 5.92 | 94.5 | 418 | 776 | 1140 |
| 94 | 0 | 0 | 0.133 | 0.358 | 0.541 | 3.77 | 72.5 | 319 | 703 | 1060 |

At every percentage value, there is a clear increase in illuminated area with respect to elevation above the surface. Growth appears generally linear but increases at a faster rate for elevations above 100m as opposed to below 100m.

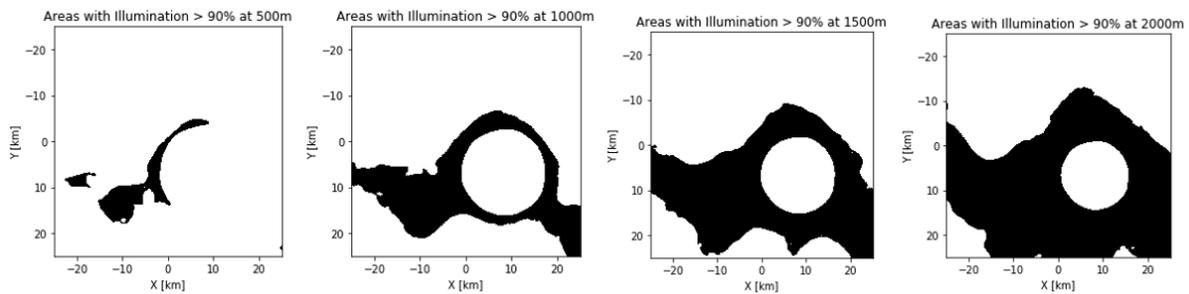

Fig 6: >90% illuminated area at 500, 1000, 1500, and 2000m.

To better quantify the growth in area with elevation, the slopes (in units of square kilometers gained per meter of elevation above the lunar surface) of the best fit lines for each percentage illumination were calculated and graphed against corresponding percentage illumination. Three slopes were calculated: one for elevations from 0-100m, one for elevations from 100-2000m, and an average of the two. Figure 7 shows that, as predicted, for the same percentage illuminations, area grows at a faster rate for higher elevations. However, the rate of increase in area decreases as illumination percentage increases.

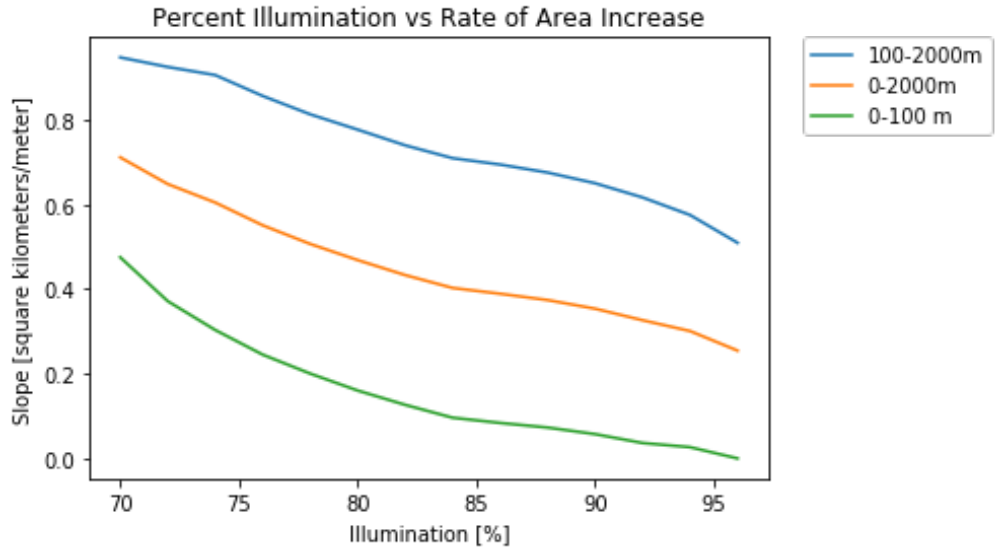

Fig 7: Slopes predicting area in square km gained per 1 m elevation for illumination percentage thresholds.

## 3.2 Illuminated length versus time of lunar day

The large high illumination areas available for tall towers (Figure 6) provide freedom to place towers in many places. However, these areas do not translate into more power availability. With almost horizontal illumination only the projected length of the regions matters.

To understand how angle of illumination causes variations in illuminated length across the lunar south pole, the illumination maps were rotated and scaled through all integer angles from 0 to 360 degrees using the ndimage library's `rotate` function from the SciPy Python package [https://docs.scipy.org/doc/scipy/reference/generated/scipy.ndimage.rotate.html]. Figure 8 shows two examples of these rotations for the 100 m map.

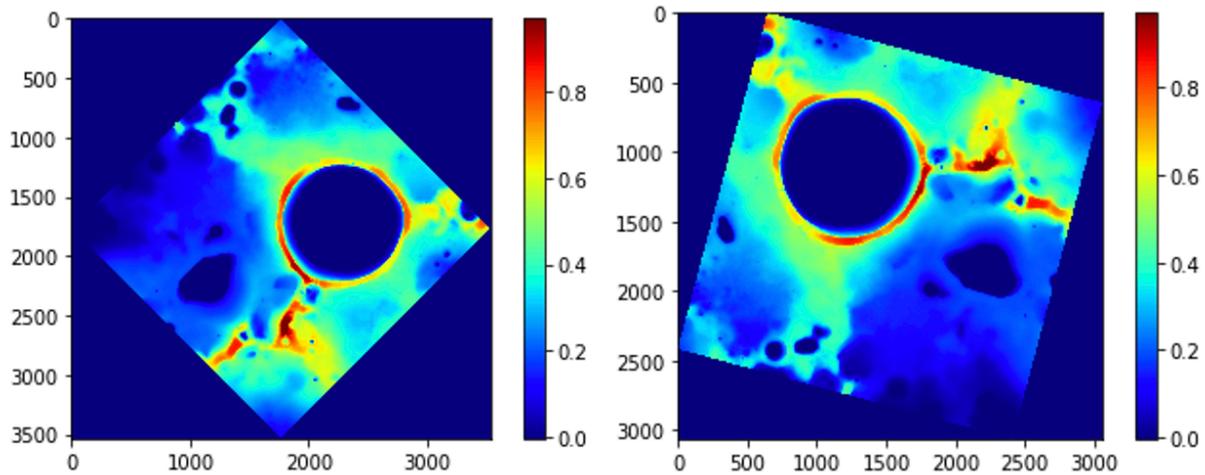

Fig 8: a selection of rotated 100m maps, left rotated at 45 degrees, right rotated at 165 degrees.

At each rotation value, the map was swept from top to bottom, aggregating the total number of illumination threshold-passing pixels per column. The total number of columns with at least one illumination threshold exceeding pixel was then aggregated and multiplied by the pixel size of 0.02 km to find the illuminated length at the current angle of illumination, which can be described as an orthographic projection to a rotating vertical plane. Note that this pixel size remains close to constant despite the map being rotated because the maps pixel dimensions increase to the orthogonal length from the top corner of the rotation to the bottom corner (for example, the 45° rotated map has dimensions 3536 x 3536 pixels). Figure 9 shows illuminated length as a function of angle of illumination for four threshold illumination percentages: >70%, >80%, >90%, and >95%. All angles from 0°-360° are shown, although it is worth noting that illuminated length values repeat after 180° (i.e., illuminated length at 0° and 180°, 90° and 270°, etc. are the same). This is because the number of illuminated pixels and columns of illuminated pixels reads the same from the top of the map down as from the bottom of the map up.

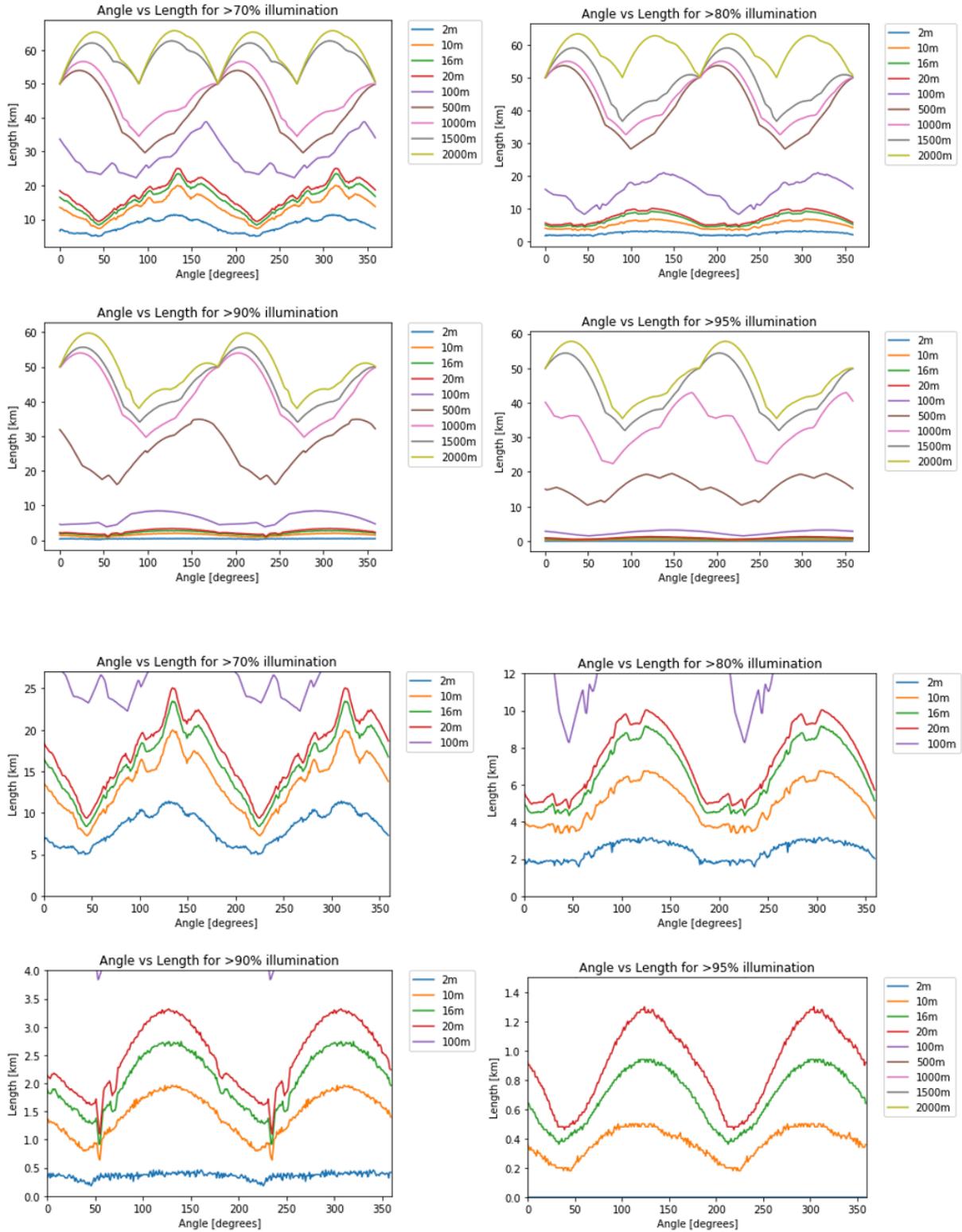

Fig 9: Illumination Angle vs Illuminated Length at Various Elevations. Top four panels: for >70%, >80%, >90% and >95% illumination; Lower four panels for heights 2, 10, 16, 20 m.

To better quantify the variation in illuminated length with respect to angle of illumination, average illuminated length over the lunar month as well as maximum & minimum illuminated lengths (all lengths in kilometers) and their corresponding angles at threshold illumination percentages of 70%, 80%, 90%, and 95% have been isolated in Table 2. The percent of usable area, defined as the useful illuminated area (average illuminated length * 0.02 km, the width of a pixel) divided by the total illuminated area and multiplied by 100, has also been included. For low towers (<100m) the usable area is substantial, ~5% or larger. For tall towers (100m plus), the usable area becomes just ~0.1% - ~1% of the total.

Table 2: Illuminated length vs. percent illumination for various heights.

| Height [m] | % Illum | Avg Len [km] | Min Len [km] | Min Angle [deg] | Max Len [km] | Max Angle [deg] | Max/Min Ratio | Usable area |
|---|---|---|---|---|---|---|---|---|
| 2 | 70 | 8.19 | 4.98 | 43, 223 | 11.4 | 130, 310 | 2.29 | 21.4% |
| 2 | 80 | 2.44 | 1.58 | 56, 236 | 3.16 | 122, 302 | 2 | 26.4% |
| 2 | 90 | 0.36 | 0.18 | 46, 226 | 0.46 | 11, 117, 157, 161, 291, 297, 337, 341 | 2.56 | 47.4% |
| 2 | 95 | 0 | 0 | - | 0 | - | 1 | - |
| 10 | 70 | 13.59 | 7.24 | 45, 225 | 19.98 | 134, 314 | 2.7 | 6.1% |
| 10 | 80 | 5.02 | 3.38 | 37, 217 | 6.76 | 123, 303 | 2 | 7.1% |
| 10 | 90 | 1.45 | 0.64 | 55, 235 | 1.96 | 123, 130, 131, 135, 303, 310, 311, 315 | 3.06 | 6.6% |
| 10 | 95 | 0.37 | 0.18 | 38, 42, 44, 45, 46, 218, 222, 224, 225, 226 | 0.5 | 96, 104, 108, 109, 111, 112, 114, 115, 116, 118, 121, 122, 123, 124, 129, 130, 131, 134, 276, 284, 288, 289, 291, 292, 294, 295, 296, 298, 301, 302, 303, 304, 309, 310, 311, 314 | 2.77 | 18.7% |
| 16 | 70 | 15.86 | 8.36 | 44, 224 | 23.46 | 134, 314 | 2.81 | 4.2% |
| 16 | 80 | 6.62 | 4.34 | 46, 226 | 9.16 | 124-125, 304-305 | 2.11 | 5.4% |
| 16 | 90 | 2.07 | 0.92 | 54-55, 234-235 | 2.74 | 127, 140, 307, 320 | 2.98 | 5.0% |
| 16 | 95 | 0.68 | 0.36 | 33, 213 | 0.94 | - | 2.61 | 5.9% |
| 20 | 70 | 17.3 | 9.38 | 44-45, 224-225 | 25.06 | 134, 314 | 2.67 | 3.6% |
| 20 | 80 | 7.32 | 4.72 | 46, 226 | 10.04 | 125, 305 | 2.13 | 4.6% |
| 20 | 90 | 2.48 | 1.1 | 55, 235 | 3.32 | 127, 307 | 3.02 | 4.3% |
| 20 | 95 | 0.91 | 0.46 | 39, 219 | 1.3 | 124, 304 | 2.83 | 4.9% |
| 100 | 70 | 28.49 | 22.26 | 87, 267 | 38.8 | 166-167, 346-347 | 1.74 | 0.95% |
| 100 | 80 | 15.36 | 8.38 | 46, 226 | 20.92 | 138, 318 | 2.53 | 1.4% |
| 100 | 90 | 6.26 | 3.84 | 53, 233 | 8.44 | 110, 290 | 2.2 | 1.6% |
| 100 | 95 | 2.5 | 1.5 | 87, 267 | 3.2 | 144-145, 324-325 | 2.13 | 1.4% |
| 500 | 70 | 42.34 | 29.66 | 97, 277 | 53.96 | 21-24, 201-204 | 1.82 | 0.19% |
| 500 | 80 | 41.40 | 28.12 | 100, 280 | 53.64 | 21, 201 | 1.9 | 0.25% |
| 500 | 90 | 26.13 | 16.04 | 65, 245 | 24.88 | 158-159, 338-339 | 2.17 | 0.44% |
| 500 | 95 | 15.63 | 10.44 | 49, 229 | 19.52 | 148, 328 | 1.87 | 0.50% |
| 1000 | 70 | 46.27 | 34.50 | 90, 270 | 56.62 | 26, 206 | 1.64 | 0.11% |
| 1000 | 80 | 44.17 | 32.56 | 93, 273 | 54.96 | 24, 204 | 1.69 | 0.13% |
| 1000 | 90 | 42.28 | 29.7 | 97, 277 | 54 | 23, 203 | 1.82 | 0.17% |
| 1000 | 95 | 33.14 | 22.38 | 79, 259 | 42.98 | 171, 351 | 1.92 | 0.24% |
| 1500 | 70 | 57.91 | 50 | 0, 90, 180, 270 | 62.7 | 127, 307 | 1.25 | 0.086% |
| 1500 | 80 | 48.56 | 36.64 | 90, 270 | 59.02 | 32, 312 | 1.61 | 0.089% |
| 1500 | 90 | 45.30 | 34.06 | 91, 271 | 55.68 | 26-27, 206-207 | 1.63 | 0.11% |
| 1500 | 95 | 43.44 | 32 | 93, 273 | 54.4 | 23-24, 203-204 | 1.7 | 0.13% |
| 2000 | 70 | 60.03 | 50 | 0, 90, 180, 270 | 65.72 | 131, 311 | 1.31 | 0.064% |
| 2000 | 80 | 58.27 | 50 | 0, 90, 180, 270 | 63.32 | 37-39, 217-219 | 1.27 | 0.077% |
| 2000 | 90 | 49.42 | 38 | 90, 270 | 59.74 | 32-24, 212-214 | 1.57 | 0.084% |
| 2000 | 95 | 47.04 | 35.45 | 90, 270 | 57.78 | 29-31, 209-211 | 1.63 | 0.094% |

### 3.2.1 Overshadowing

Not all of the high illumination areas will necessarily be useful for power generation. As the Sun is always near the horizon, overshadowing may occur when an illuminated region on a rotated map lies behind another illuminated region on a rotated map. For example, the two regions on opposite sides of the rim of Shackleton crater rarely can both see the Sun without overshadowing. More generally, from the rotated maps in Fig. 8, this can be visualized as a red-orange region that occurs closer to the bottom of the map than a red-orange region in the same column of the map but occurring closer to the top. To calculate overshadowing, the rotated maps were swept for illuminated *clusters* rather than individual illuminated pixels. This is because there are extremely few, if any, instances of single illuminated pixels in a column. A sweep for consecutive illuminated pixels would therefore indicate that virtually the entire illuminated length is overshadowed, which is not the case in practice. Illuminated clusters, in contrast, are defined as areas that begin with a threshold-passing illuminated pixel and end when a pixel's illumination value dips below a separate threshold illumination value (60% illumination at <=100m, and 80% illumination above that). By sweeping for illuminated clusters, overshadowing is only accounted for when there are large swatches of area that will not receive illumination due to a previously illuminated area.

Overshadowing applies more to lower heights (<500m) than higher heights. As shown by the last column in Table 2 describing usable area, as height increases, the percentage of usable area decreases. This is because for higher heights, such a high percentage of total map area is illuminated that longer illuminated lengths constitute increasingly smaller fractions of the total illuminated area. Within each height map, percent usable area increases as illumination percentage increases for similar reasons - at higher illumination percentages, illuminated area becomes sparser, and comparatively smaller decreases in illuminated length make for larger fractions of usable area compared to larger decreases in illuminated area. At heights over 100m, overshadowing is present at all times of the lunar day, so the optimization of tower placement might have more readily found solutions. At heights 100 m and below, overshadowing makes a much clearer difference as the illuminated area is restricted to a narrow swath of crater rim or the connecting ridge between Shackleton and de Gerlache craters.

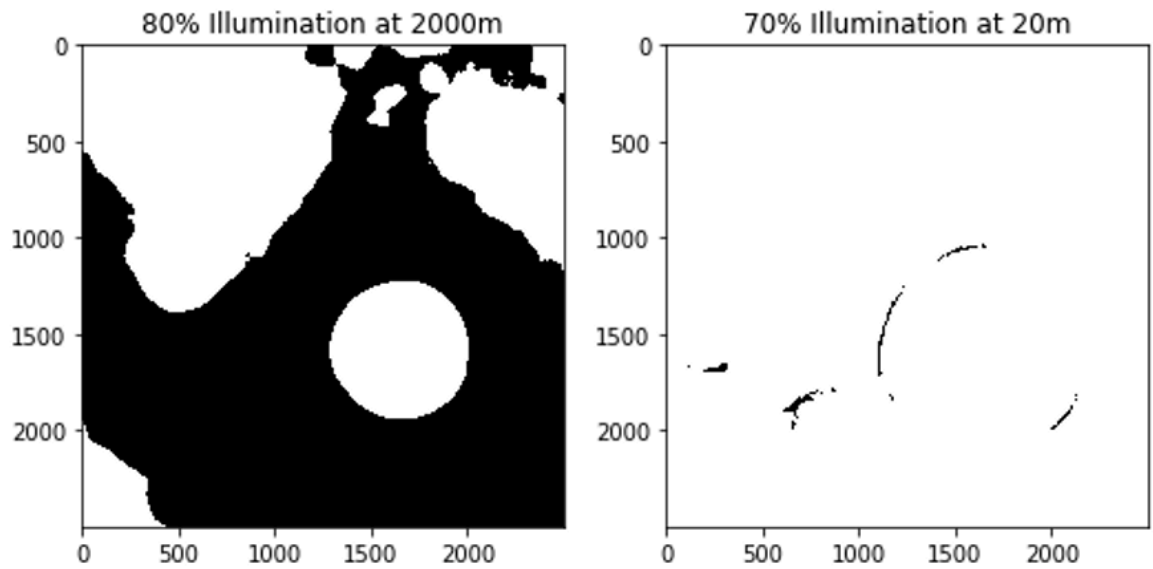

Figure 10: Maps demonstrating high amounts of overshadowing at highly illuminated heights and high amounts of overshadowing and sparsely illuminated heights.

After the rotated maps were swept for illuminated clusters, the total number of illuminated clusters per column (generally ranging from 0-3) was aggregated, and then the total number of columns with more than one illuminated cluster (which indicates overshadowing) was aggregated and multiplied by 0.02 km for the total overshadowed length. Figure 10 shows illuminated and overshadowed length plotted together as a function of angle of illumination for the previous four threshold illumination percentages. Overshadowing goes from a small fraction of the total length (see e.g. 2m) to almost 100% (see e.g. 500m, 70%), depending on height and illumination threshold.

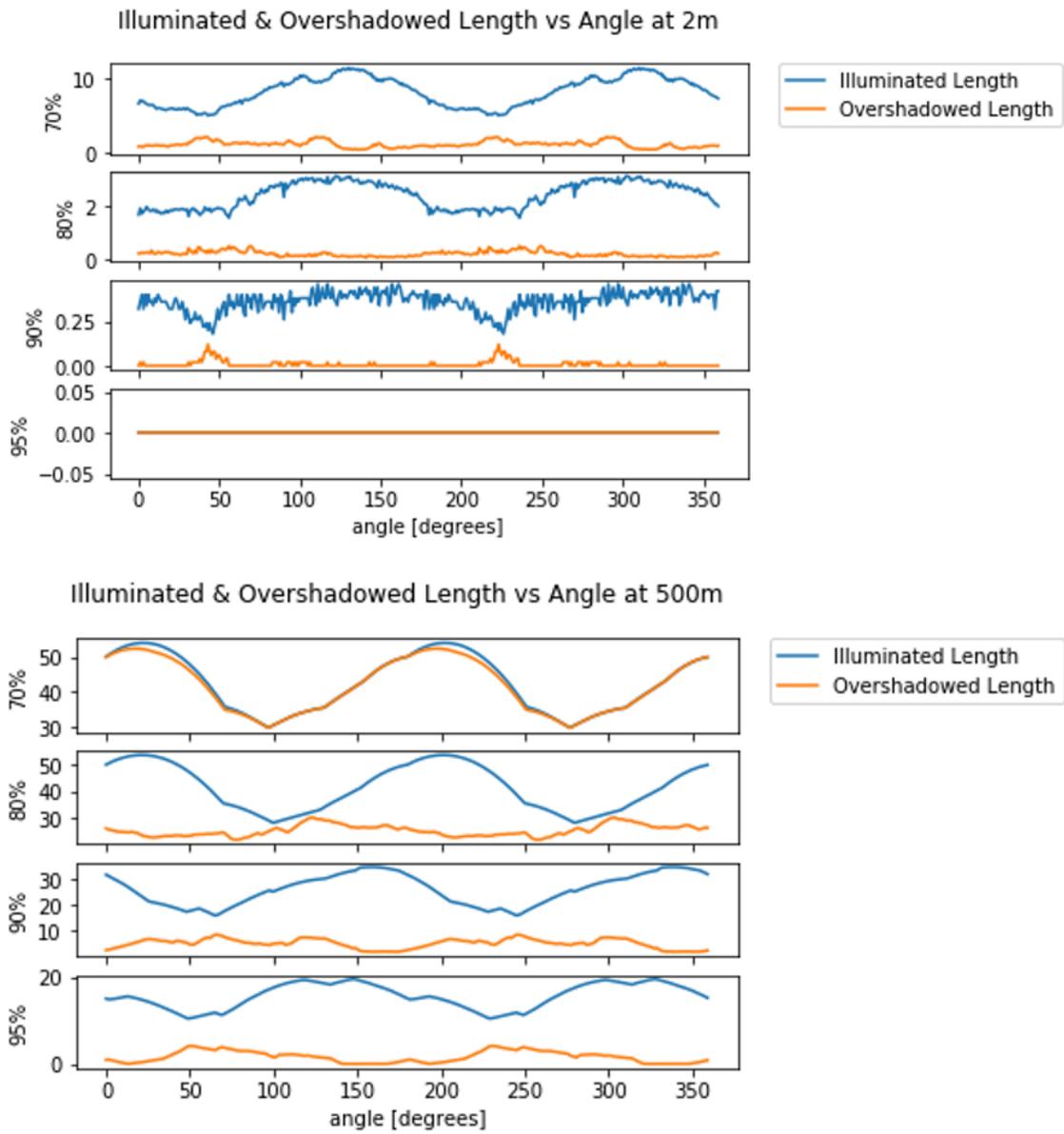

Fig 11: Illumination Angle vs Illuminated & Overshadowed Length at 2m and 500m

## 4. Discussion

### 4.1 Illuminated Area, Average Illumination, and Elevation

Figures 1 and 2 show illumination percentages at various elevations. Areas suitable for building solar power towers are generally areas with >80% illumination (marked by red and dark red areas on the maps). Solar power towers placed in these areas will therefore be able to generate solar power greater than 80% of the time. Gläser et al. [8] showed that the times without Sun are short, of order a few days, so at these locations relatively modest storage capacity will be needed to provide continuous power.

Figure 1 shows that high illumination areas are rare at lower elevations (<100m). However, the illumination maps for 100m and higher show much more high illumination areas (Figure 2). The total area available for a range of threshold illumination values is shown in Figure 4. At >80% illumination, elevations from 0-20m have less than 3.5 square km, while at 100m of elevation, there is 22 square km of highly illuminated area. Solar power towers built to the height of at least 100m, and preferably higher, will likely be able to generate significant almost-continuous solar power.

Figure 4 shows that for higher illumination percentages, smaller increases in illuminated area occur as elevation increases. This can be seen in the slopes of the area-elevation lines shown in Figure 7. They show explicitly that, as illumination percentage increases, the rate at which illuminated area increases with elevation decreases. Figure 5 predicts that about a 1.33 meter increase in elevation will correspond with a 1 square km increase in >80% illuminated area, while a 2.33 meter elevation increase is required to gain 1 square km in >96% illuminated area.

**4.2 Projected Illuminated Length versus Monthly Phase**

Over the course of the lunar day, the Sun moves around the horizon at the poles, illuminating them from different angles. Viewed from the Sun, the projected length of terrain illuminated changes, hence the amount of power that can be generated changes. Although the illumination maps indicate the average illumination of each pixel, or 20m by 20m square, on the map, if any of these squares are overshadowed by previous illuminated points, they will not increase the overall illuminated length. This turns out to be a common effect.

Figure 9 shows that illuminated length cycles twice over the course of a month, reaching two peaks and two valleys. The peaks represent the longest available illuminated length, and the valleys represent the shortest. Table 2 gives the maximum to minimum ratio for each case. These values range from 1.25 to 3.06, and trend lower for taller towers. Those values closer to 1 indicate a roughly constant amount of illumination throughout the lunar day, such as the 80% illumination map at 2000m, while higher values indicate larger differences in length depending on angle of rotation, such as the 90% illumination map at 10 m, as the maps in Figure 10 in the overshadowing section demonstrate.

**4.3 Available Power**

The projected illuminated lengths for each elevation at each phase during a month can be used to make a first order determination of the maximum, minimum, and average solar power output over the course of a lunar month.

To make these estimates, we first need a value for solar irradiance, or the power per square meter supplied by the Sun at a distance of 1 AU. This value is 1366.1 W/m$^2$, which is equivalent to 1.37 kW/m$^2$ [15].

We also need the power per square meter provided by space-qualified solar panels in cis-lunar space. The average efficiency for present day silicon solar cells is ~15%, as presented by Gibb [16] and Hoang [17]. Present day best case space solar panels can achieve efficiencies nearly double that, such as solar panels from Spectrolab [18] and Azurspace [19] with 28%-32% efficiency. However, these solar panels have much too small surface areas for our purposes. Therefore, we estimate the power per square meter available using present day space solar cells to be 0.21 kW/m$^2$.

The average, minimum, and maximum illuminated lengths in meters, at each elevation and each percentage threshold can be multiplied by the elevation to get an illuminated area; the power per square meter then gives an estimate of the maximum power generated for the average, minimum, and maximum lengths along the lunar surface (termed average power, minimum power, and maximum power). This calculation assumes that solar panels are placed along the entire available length and up the entire height. For example, at 2 m, the average illuminated length is

8.19 km. Our naïve calculation then assumes that we have 2 * 8190 = 16280 m² of solar panel area. This is furthermore an approximate calculation because the change in illumination with location is not taken into account. Nevertheless, this approach provides a first estimate that sets the scale of the resource.

With increases in height, the illuminated lengths increase, so the power was estimated in steps of elevation. To calculate a lower limit to the power generated, the elevation step size was multiplied by the illuminated length at the *lower* elevation before being multiplied by the power per square meter before finally being converted to megawatts and added to a cumulative total. For example, to determine a lower limit to the average power generated at 500 m at 70% illumination, the elevation step size (500m - 100m) was multiplied by the average illuminated length at 100 m (Table 2) and the power generated per square meter (0.22 kW/m²) before finally being added to the lower power limit at the 70% entry at 100 m. Similarly, to calculate an upper limit to the power generated, the elevation step size was instead multiplied by the illuminated length at the *higher* elevation before being multiplied by the power per square meter, before being converted to megawatts and added to a cumulative total. At 2 m, the underestimates were calculated using the rotational data at 0 m, which is not shown here because presumably no towers will be built at ground level. The mean (which functions as a linear interpolation of the lower and upper estimates) and results of both estimates are given in Table 3 in the form "mean (lower estimate - upper estimate)". The means of the lower and upper limits for average power are graphed in Figure 12 (with the exception of 2 m, where the ranges between maximum and minimum power are too large for the mean to be meaningful).

Table 3: Mean, lower, and upper limits to the average, minimum, and maximum power for various heights and threshold percent illuminations in the form "mean (lower - upper)".

| Height [m] | Illumination [%] | Time-Averaged Power [MW] | Min Power [MW] | Max Power [MW] |
|---|---|---|---|---|
| 2 | 70 | 2.22 (0.83 - 3.6) | 1.39 (0.58 - 2.19) | 3.12 (1.21 - 5.02) |
| 2 | 80 | 0.56 (0.05 - 1.07) | 0.36 (0.02 - 0.7) | 0.735 (0.08 - 1.39) |
| 2 | 90 | 0.08 (0 - 0.16) | 0.04 (0 - 0.08) | 0.1 (0 - 0.2) |
| 2 | 95 | 0 (0 - 0) | 0 (0 - 0) | 0 (0 - 0) |
| 10 | 70 | 21.4 (15.2 - 27.5) | 12.1 (9.34 - 14.9) | 30.7 (21.3 - 40.2) |
| 10 | 80 | 7.13 (4.34 - 9.91) | 4.73 (2.8 - 6.65) | 9.47 (5.64 - 13.3) |
| 10 | 90 | 1.67 (0.63 - 2.71) | 0.765 (0.32 - 1.21) | 2.23 (0.81 - 3.65) |
| 10 | 95 | 0.325 (0 - 0.65) | 0.16 (0 - 0.32) | 0.44 (0 - 0.88) |
| 16 | 70 | 40.8 (33.2 - 48.5) | 22.4 (18.9 – 26.0) | 59.4 (47.6 - 71.2) |
| 16 | 80 | 14.8 (11.0 - 18.7) | 9.82 (7.26 - 12.4) | 20.0 (14.6 - 25.4) |
| 16 | 90 | 3.99 (2.54 - 5.44) | 1.79 (1.16 - 2.42) | 5.34 (3.4 - 7.27) |
| 16 | 95 | 1.02 (0.49 - 1.55) | 0.52 (0.24 - 0.8) | 1.39 (0.66 - 2.12) |
| 20 | 70 | 55.4 (47.1 - 63.7) | 30.2 (26.3 - 34.2) | 80.7 (68.3 - 93.2) |
| 20 | 80 | 21.0 (16.8 - 25.1) | 13.8 (11.1 - 16.5) | 28.4 (22.6 - 34.2) |
| 20 | 90 | 5.99 (4.36 - 7.62) | 2.68 (1.97 - 3.39) | 8 (5.81 - 10.2) |
| 20 | 95 | 1.72 (1.09 - 2.35) | 0.88 (0.56 - 1.2) | 2.38 (1.49 - 3.26) |
| 100 | 70 | 458 (352 - 565) | 309 (191 - 426) | 643 (509 - 776) |
| 100 | 80 | 221 (146 - 295) | 129 (94.2 - 164) | 301 (199 - 402) |
| 100 | 90 | 82.9 (48.0 - 118) | 46.2 (21.3 – 71.0) | 111 (64.2 - 159) |
| 100 | 95 | 31.7 (17.1 - 46.4) | 18.1 (8.66 - 27.6) | 42.0 (24.4 - 59.6) |
| 500 | 70 | 3570 (2860 - 4290) | 2590 (2150 - 3040) | 4720 (3920 - 5520) |
| 500 | 80 | 2720 (1500 - 3940) | 1740 (832 - 2640) | 3580 (2040 - 5120) |
| 500 | 90 | 1510 (599 - 2420) | 921 (359 - 1480) | 1580 (807 - 2350) |
| 500 | 95 | 829 (237 - 1420) | 543 (141 - 946) | 1040 (306 - 1780) |
| 1000 | 70 | 8450 (7520 - 9380) | 6120 (5410 - 6830) | 10800 (9860 - 11800) |
| 1000 | 80 | 7420 (6050 - 8800) | 5070 (3920 - 6220) | 9550 (7940 - 11200) |
| 1000 | 90 | 5270 (3470 - 7070) | 3440 (2120 - 4750) | 5920 (3540 - 8290) |
| 1000 | 95 | 3510 (1960 - 5070) | 2350 (1290 - 3410) | 4480 (2450 - 6510) |
| 1500 | 70 | 14200 (12600 - 15800) | 10800 (9210 - 12300) | 17400 (16100 - 18600) |
| 1500 | 80 | 12500 (10900 - 14100) | 8880 (7510 - 10300) | 15800 (14000 - 17700) |
| 1500 | 90 | 10100 (8120 - 12100) | 6940 (5390 - 8500) | 11900 (9480 - 14400) |
| 1500 | 95 | 7720 (5600 - 9850) | 5340 (3750 - 6930) | 9840 (7180 - 12500) |
| 2000 | 70 | 20700 (19000 - 22400) | 16300 (14700 - 17800) | 24400 (23000 - 25900) |
| 2000 | 80 | 18400 (16300 - 20500) | 13600 (11500 - 15800) | 22600 (20500 - 24600) |
| 2000 | 90 | 15300 (13100 - 17500) | 10900 (9140 - 12700) | 18300 (15600 - 21000) |
| 2000 | 95 | 12700 (10400 - 15000) | 9050 (7270 - 10800) | 16000 (13200 - 18800) |

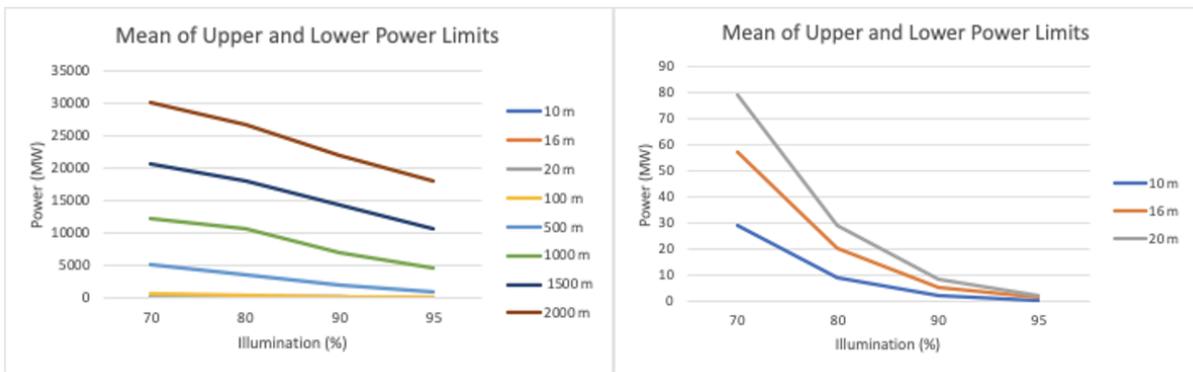

Figure 12: Mean of upper and lower power limits plotted by height, with illumination percentage on x axis and power in megawatts on y axis. Right: blowup panel for lower heights (10-20m).

For near-term towers up to 20m the available time-averaged power ranges from ~2 at 2m and ~55 MW at 20m for 70% illumination, down to ~0.1 MW at 2 m to ~6 MW at 20 m for 90% illumination. Instead for tall towers, 100 m and up, ~450 - ~20000 MW are available for 70% illumination, and ~82 MW to ~15000 MW for 90% illumination (Table 3).

**5. Discussion**

We have found that the maximum power available from solar panels at the Peaks of Eternal Light is substantial and depends strongly on the height of the panels above the local surface. The average power is in the few megawatt range for shorter towers (<= 100 m), and the few gigawatt range for taller towers (500+ m). The variations with time of the lunar day are quite large (factors of 1.1- 3, Table 4).

Are the power capabilities listed in Table 3 appropriate for lunar mining? Sowers and Dreyer [6] use a "thermal mining" approach and find that 0.5 MW is sufficient to sustain production of hydrogen and oxygen from water at a rate of 1600 mt/year. Towers of around 16 m in height are needed to provide this level of power for >90% of the time over a full lunar day. In contrast, Karnuta et al. [7] estimate that extraction of 2450 tons/year of water from the permanently dark regions would require much higher power levels of 400 - 1400 MW, with the higher value applying to lower concentrations of ice in the regolith. These power needs require towers somewhat above 500 m for 90% availability throughout the lunar day. Understanding which power range will be required to create a successful lunar ice mining operation is necessary to deciding the scale of effort needed to provide this power, and so the feasibility of the project.

There are a number of limitations to the present analysis:

1. The differences between the upper and lower limits derived above are substantial. Table 4 and Figure 13 show the spread of their ratio, with a mean between 2.4 - 3. The vast majority of illumination and height values have a ratio under 2, but the mean is skewed high by a few outliers. To determine more accurate values would require the generation of many more illumination maps for intermediate heights, which is too computationally demanding for this preliminary study.

Table 4: Ratio of upper limit to lower limit of average, minimum, and maximum power for various heights and threshold percent illuminations.

| Height [m] | Illumination [%] | Avg Ratio | Min Ratio | Max Ratio |
|---|---|---|---|---|
| 2 | 70 | 4.34 | 3.78 | 4.15 |
| 2 | 80 | 21.4 | 35 | 17.38 |
| 2 | 90 | N/A | N/A | N/A |
| 2 | 95 | N/A | N/A | N/A |
| 10 | 70 | 1.81 | 1.6 | 1.89 |
| 10 | 80 | 2.29 | 2.38 | 2.36 |
| 10 | 90 | 4.31 | 3.79 | 4.51 |
| 10 | 95 | N/A | N/A | N/A |
| 16 | 70 | 1.47 | 1.38 | 1.5 |
| 16 | 80 | 1.71 | 1.71 | 1.75 |
| 16 | 90 | 2.15 | 2.09 | 2.14 |
| 16 | 95 | 3.17 | 3.34 | 3.22 |
| 20 | 70 | 1.36 | 1.31 | 1.37 |
| 20 | 80 | 1.5 | 1.5 | 1.52 |
| 20 | 90 | 1.75 | 1.73 | 1.76 |
| 20 | 95 | 2.16 | 2.15 | 2.19 |
| 100 | 70 | 1.61 | 2.23 | 1.53 |
| 100 | 80 | 2.03 | 1.75 | 2.02 |
| 100 | 90 | 2.46 | 3.33 | 2.48 |
| 100 | 95 | 2.71 | 3.19 | 2.45 |
| 500 | 70 | 1.51 | 1.42 | 1.41 |
| 500 | 80 | 2.64 | 3.18 | 2.52 |
| 500 | 90 | 4.04 | 4.13 | 2.91 |
| 500 | 95 | 6 | 6.73 | 5.81 |
| 1000 | 70 | 1.25 | 1.27 | 1.2 |
| 1000 | 80 | 1.46 | 1.59 | 1.41 |
| 1000 | 90 | 2.04 | 2.24 | 2.34 |
| 1000 | 95 | 2.6 | 2.65 | 2.66 |
| 1500 | 70 | 1.25 | 1.34 | 1.16 |
| 1500 | 80 | 1.3 | 1.37 | 1.27 |
| 1500 | 90 | 1.49 | 1.58 | 1.52 |
| 1500 | 95 | 1.76 | 1.85 | 1.74 |
| 2000 | 70 | 1.18 | 1.22 | 1.13 |
| 2000 | 80 | 1.27 | 1.37 | 1.21 |
| 2000 | 90 | 1.34 | 1.39 | 1.35 |
| 2000 | 95 | 1.45 | 1.49 | 1.44 |

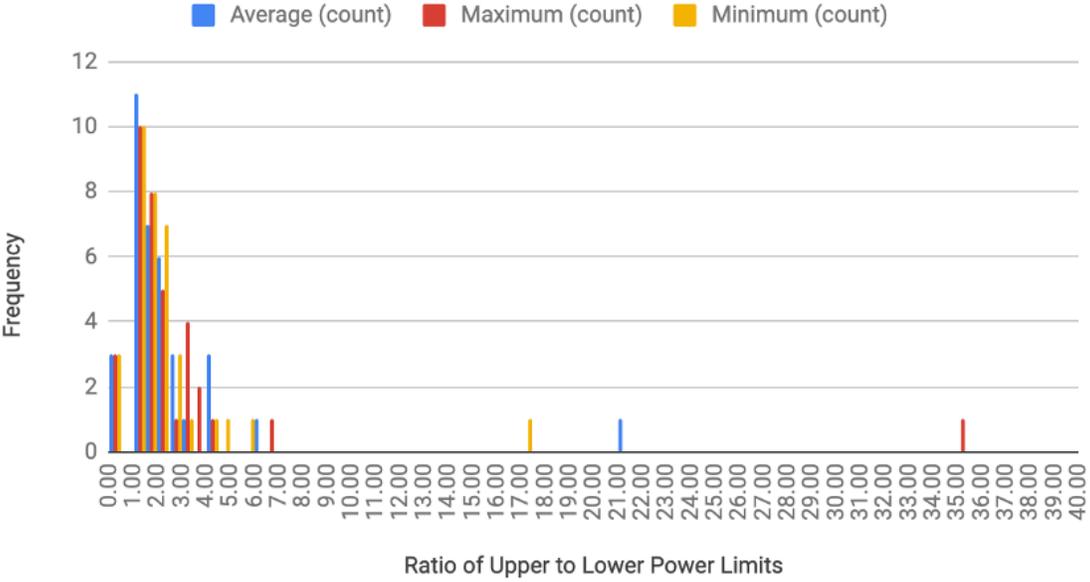

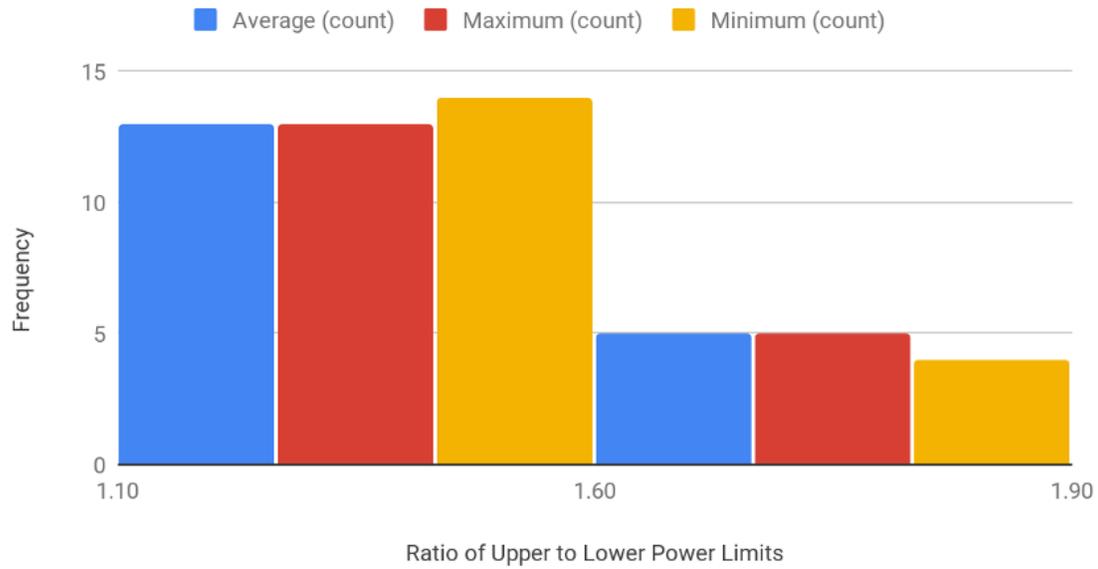

Figure 13: Histograms of average, maximum, and minimum ratios of upper to lower power limits, bottom histogram limited to ratios under 2.

2. Similarly, the true variation of illumination through the lunar day should use the illumination map for each time of day, rather than a single 18.6 year average illumination data as used here. This also overlooks the seasonal illumination variation on the lunar surface, which includes summer periods with as much as 100% illumination as well as winter periods with 0% illumination. This is computationally demanding and was beyond the scope of this proof-of-principle study. Follow-up work should investigate how important this approximation is to the available power.

3. Another consideration is that illumination percentages sometimes vary greatly from pixel to pixel - in the 70% and higher range, some neighboring pixels vary only by 1% (i.e. e.g. 78% to 79%), while others may vary as much as from 25% (70% to 95%). In addition, this variation could mean that the map is under-sampled at 20 m resolution. There is potentially less overshadowing than expected. For example, if two pixels both have 70% illumination, at the most extreme, one pixel could be illuminated for the first 70% of the lunar month, while the other could be illuminated for the second 70% of the lunar month, giving the two pixels only 40% of overlapping, overshadowed solar illumination and 60% of un-shadowed illumination otherwise. To take this into account, more detailed maps of hourly illumination over the course of a lunar day at various elevations would be necessary.

4. This naïve overshadowing estimate also ignores the curvature of the lunar surface. Due to the curvature of the Moon (1737.1 km radius), an object on a perfectly spherical Moon would drop 720 m over a 50 km range by the Pythagorean theorem. Therefore, two theoretical solar arrays along the same Sunline separated by 50 km would have zero overshadowing under a height of 720 m. This is computationally demanding and beyond the scope of this paper, but will be necessary to assess in more detailed studies.

5. Our power generation calculation does not include a temperature estimate. This estimate would require a thermal balance, which would include factors sunlight, albedo, infrared, and surface emissivity and absorptivity. This is outside the scope of our simple calculation and requires more details about the specific solar cells utilized for the project, which must be tested and validated with regard to structure and temperature for lunar surface use. The NASA Lunar Vertical Solar Array Technology (Lunar VSAT) project is currently embarking on this endeavor [20].

Note that we have assumed we can cover the entire illuminated area with solar panels. However, designing a realistic and practical arrangement for cost-effective solar power stations at the PELs will require considerable analysis. Simply covering the entire high illumination areas at the PELs with solar panel towers would be wasteful, as many would be overshadowed by other towers. A more in-depth analysis would be required to determine optimal placement of solar power towers and more exact estimates of solar power generated by that placement. One factor not considered in the current approximation is overshadowing. As shown in Fig. 10, part of the illuminated length at each angle of rotation consists of overshadowed length.

The simplest solution to this would be to place solar power towers along the longest illuminated length at every angle, with the solar panels such that they are facing radially outward. An alternative solution requiring fewer towers would involve placing the solar panels along the longest illuminated length for the most favorable angle of illumination, but not along any overshadowed portions along this length. The solar panels are then rotated to face the illumination direction. Then, place more solar panels in columns with illuminated pixels but presently without solar panels. This will decrease the number of solar panels needed because solar panels would only be placed only in non-overshadowed illuminated locations as opposed to all illuminated locations. However, because the maximum illuminated length at every angle will still be covered by solar panels (assuming the solar panels are able to rotate to face the angle of illumination), the amount of solar power generated should not decrease.

The current analysis also assumes that solar panels are only placed along pixels above the chosen percent illumination threshold and not on all available pixels at all heights. For example, for the upper bound estimate of average power output for a 100 m tall tower at 70% illumination, 28.49 km is the length along which solar panels would be placed from elevations from 20 m to 100 m. In contrast, from elevations of 0 m to 2 m, only 8.19 km of solar panels will be placed, as only these receive >70% illumination. Power generation could be increased if solar panels were placed at every available point in the tower, regardless of the illumination threshold. Future analyses should consider the trade off of benefits of increased power generation against the costs of transporting and mounting additional solar panels.

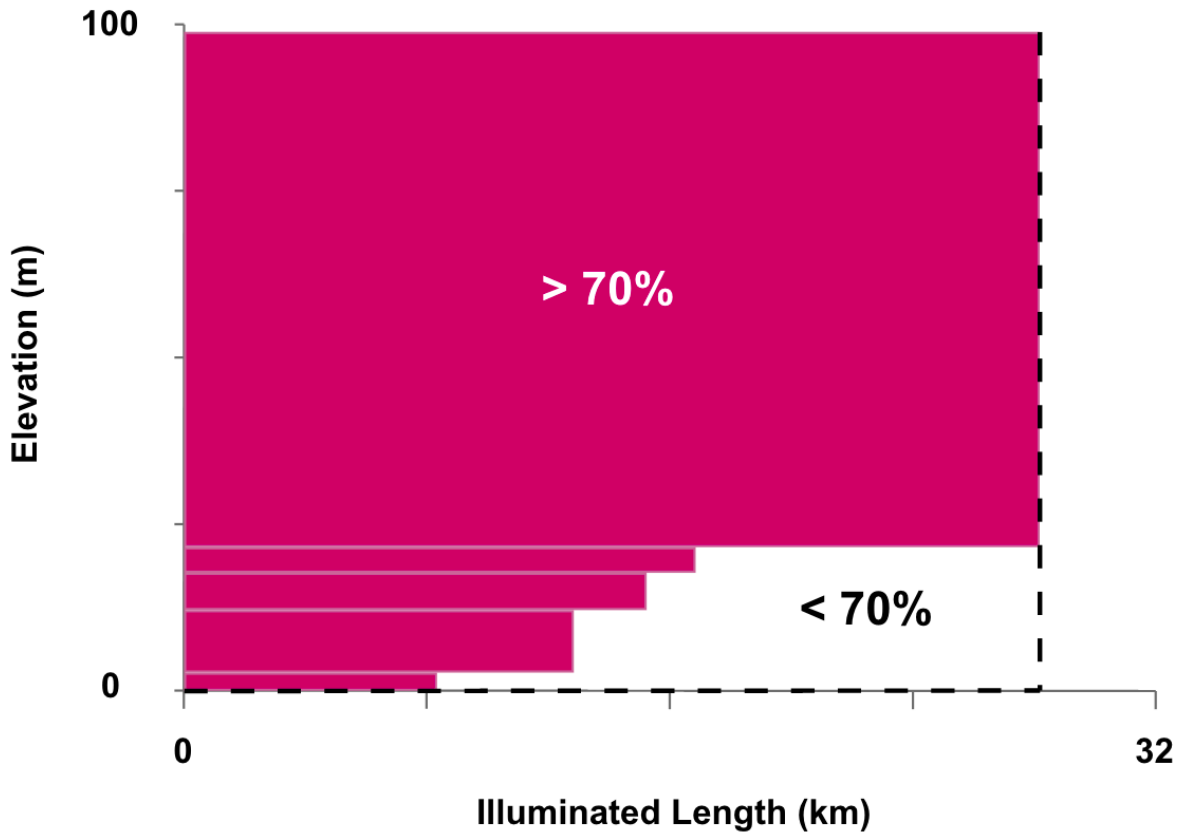

Figure 14: Diagram illustrating the approximate placement of solar panels by percent illumination vs across the total area available.

Any real arrangement will optimize the maximum power for the least number and height of panels. Halbach et al. [9] have begun this optimization process.

In addition, as mentioned in the previous section, present-day baseline solar panel technology reaches an efficiency of 15% [16]. For present-day best-case solar panels, the efficiency has doubled to ~30% [16]. Within the next decade, concentration could improve this value to ~50%.

Therefore, as solar panel technology advances and higher efficiency solar panel technology proliferates, the estimated power outputs of lunar solar power generation towers could increase by factors of 2-3.

There will also be transmission losses from the power production point to the power utilization point. In the case of microwave transmission, these are free space path losses and can be substantial.

Furthermore, high solar towers have the drawback of being candidates for erosion by blast ejecta by any landers. Pre-prepared landing pads can decrease this risk. High solar towers also may pose descent and ascent flight risks.

The cost per megawatt of solar power at the PELs will be crucial to any mining decision. We can estimate the price of the solar panels to generate the lower and upper limits of power calculated in Tables 2 and 3. Present-day terrestrial silicon solar panels cost <$1/W, while space-qualified solar panels, mostly using germanium III/IV semiconductors, cost >$100/W [16]. There are opportunities to decrease space-specific solar panel costs [16]. Thus, $10^6$/MW will serve as a lower limit, and $10^8$/MW as an upper limit on the price of the panels alone. Transport costs for the panels may be an important factor. Space qualified panels produce ~30-50 W/kg, or 20 - 30 mt/MW (Figure 7) [16]. For now, this is a large mass to land on the Moon but may be quite modest on the decade plus timescale on which lunar ice mining will need megawatts [21].

## 6. Conclusions

The scale of maximum solar power that can be generated on the Peaks of Eternal Light near the lunar south pole is of order hundreds of megawatts for panels on towers up to 100 meters in height. Such towers could very well be deployed in the near term. There are PELs beyond the 25 km radius of the south Pole studied here, notably Mount Malapert. These could be included in future studies.

Because of the geography of the PELs the maximum power attainable varies by factors up to two through a lunation, and is close to being symmetrical so that the highest value repeats roughly every 14 days.

If much taller towers could be built, up to 2 km in height, then the maximum solar power from the PELs rises to of order several gigawatts. Such towers may be buildable from local regolith (Ruppert et al. 2021) [10]. At heights above 500 m the area of the PELs increases greatly, such that the entire 50 km wide region investigated can supply solar power. Some week to week variation up to a factor of 2 is still present. The approximations made in calculating these preliminary values imply an uncertainty of a factor of 2.4-3. It is possible that these values could all be doubled by using more efficient solar panels that are now under development.

However, the near-horizon location of the Sun on the PELs means that overshadowing of panels by others is common. As a result, many high illumination locations will not be useful for solar power generation once other areas have emplaced towers. Studies of the optimal placement of towers of various heights would be valuable. These studies should include more accurate treatments of illumination versus height and location.

The power requirements for mining water from Shackleton crater and other nearby permanently dark craters are compatible with the maximum solar power available at the PELs at an extraction scale of several thousand tons per year. If this theoretical level of power turns out to be impractical, for example through transmission losses, then alternative power sources, in particular nuclear, may be required. If instead the power from the PELs is sufficient for mining, then the strong overshadowing implies that ways of developing this capability to maximize the benefit, while minimizing conflict, will be needed.

**Acknowledgments:** AR thanks the Smithsonian Astrophysical Observatory for financial support while this work was carried out. We thank Robin Wordsworth for helping this project get started. We thank James Fincannon and Dennis Wingo for helpful comments.

**Appendix**

Extension of Figure 11: illuminated and overshadowed length vs rotation angle at heights of 10 m, 16 m, 20 m, 100 m, 1000 m, 1500 m, and 2000 m.

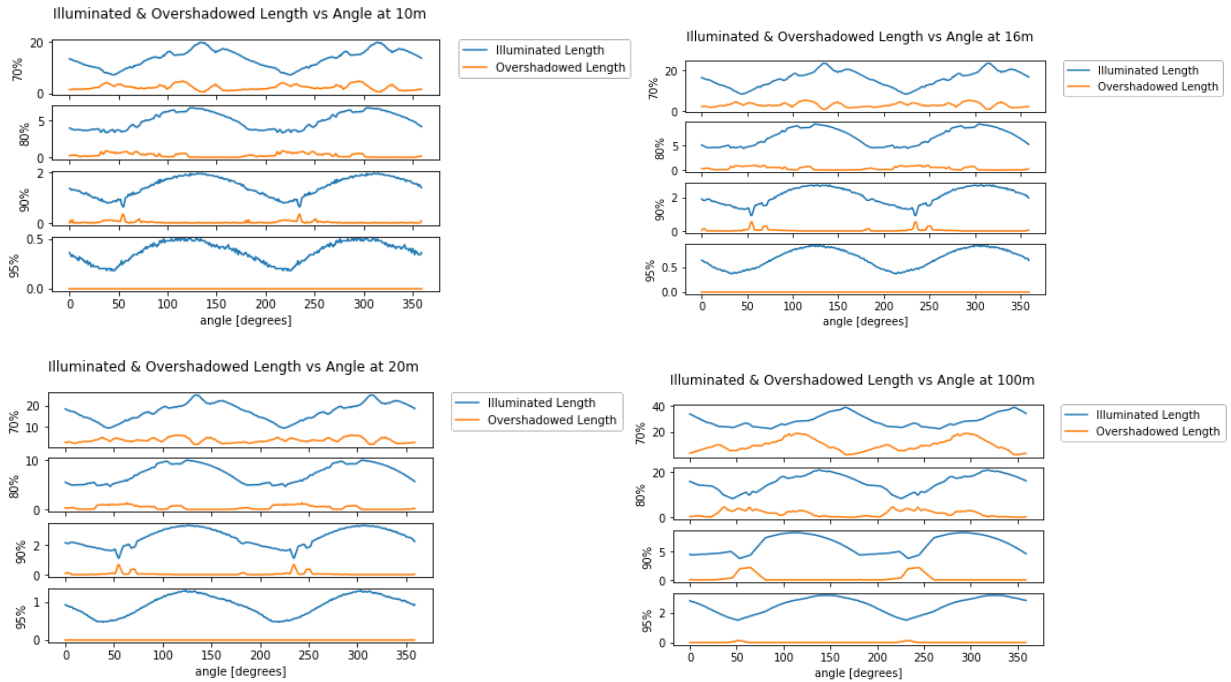

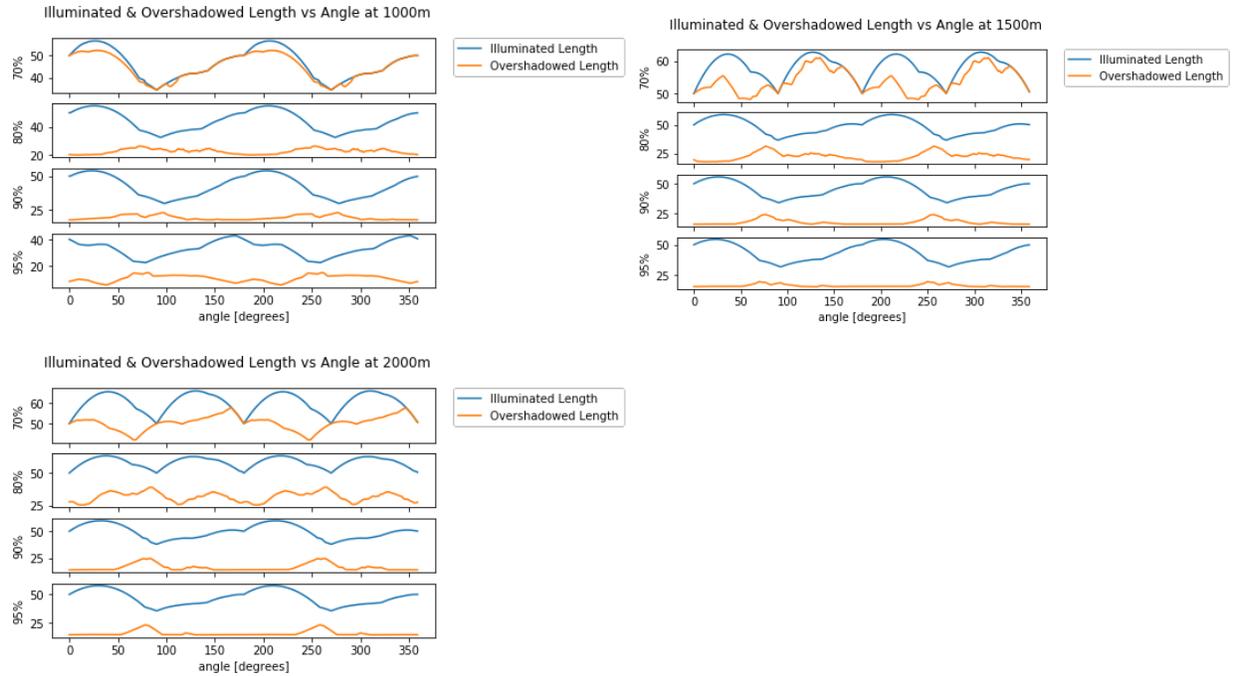

# **VITAE**

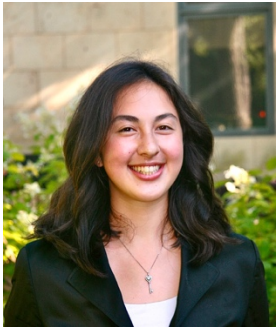

**Amia Ross:** Amia Ross is an undergraduate at Harvard University pursuing a degree in Computer Science on the Mind, Brain and Behavior track.

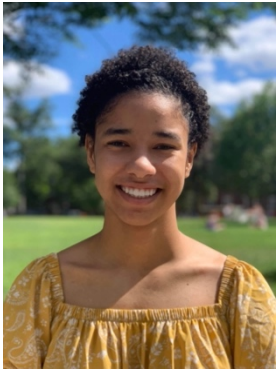

**Sephora Ruppert:** Sephora Ruppert is Harvard University undergraduate student pursuing a joint degree in Physics and Engineering Sciences.

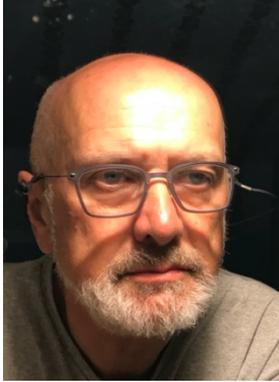

**Dr. Martin Elvis:** Dr. Martin Elvis is a highly cited astrophysicist (over 30,000 peer citations) who has published some 400 papers on supermassive black holes, seen as quasars, out to the edge of the universe over the course of many years. Lately, concerned about the growing cost of space telescopes, he has turned to researching the astronomy needed to enable asteroid mining, with a view to cutting those costs in the long run. He has published widely on issues related to asteroid mining and the space economy. He is proud that he is (probably) the first professional astronomer to visit the Harvard Business School on business. He obtained his PhD in X-ray astronomy in 1978 in the UK, and has worked at the Harvard-Smithsonian Center for Astrophysics ever since on a series of space X-ray telescopes, culminating with the Chandra X-ray Observatory. He is a fellow of the American Association for the Advancement of Science, a Member of the Aspen Center for Physics, and is past-Chair of the Hubble Space Telescope Users' Committee and of the High Energy Division of the American Astronomical Society. Asteroid 9283 Martinelvis is named after him.

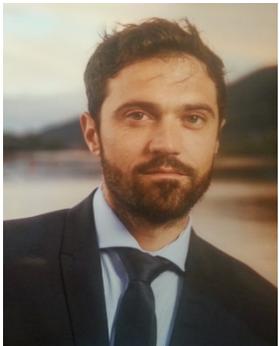

**Dr. Philipp Gläser:** Philipp Gläser is a post-doc level researcher at the Technische Universität of Berlin at the department of planetary geodesy. He researched the lunar polar environment with respect to illumination and temperature using topographic data obtained by the Lunar Orbiter Laser Altimeter. He mapped several potential landing sites near the lunar poles where optimal illumination conditions are found and permanently shadowed regions, potentially containing water ice, are in reach.